# MRADSIM (Matter-RADiation Interactions SIMulations)

**Ali Behcet Alpat[a]\***, **Giovanni Bartolini[b]**, **Talifujiang Wusimanjiang[c]**, **Haider Raheem[d]**, **Ersin Huseyinoglu[e]**, **Raziye Bayram[f]**, **Arca Bozkurt[g]**, **Deniz Dolek[h]**, **Lucia Salvi[i]**, **Ahmed Imam Shah[j]**, **Nora Ciccarella[k]**, **Yakup Bakıs[l]**, **Stefano Gigli[m]**

[a] *Istituto Nazionale di Fisica Nucleare (INFN), Sezione di Perugia, Via A.Pascoli snc, 06123, Perugia, Italy, behcet.alpat@pg.infn.it*
[b] *BEAMIDE s.r.l., Via Campo di Marte 4/o, 06124, Perugia, Italy, giovanni.bartolini@beamide.com*
[c] *Istituto Nazionale di Fisica Nucleare (INFN), Sezione di Perugia, Via A.Pascoli snc, 06123, Perugia, Italy, wusimanjiang.talifujiang@pg.infn.it*
[d] *BEAMIDE s.r.l., Via Campo di Marte 4/o, 06124, Perugia, Italy, haider.raheem@beamide.com*
[e] *IRADETS A.Ş., Teknopark İstanbul, 34906, Pendik/İstanbul, Türkiye, ersin.huseyinoglu@iradets.com*
[f] *IRADETS A.Ş., Teknopark İstanbul, 34906, Pendik/İstanbul, Türkiye, raziye.bayram@iradets.com*
[g] *IRADETS A.Ş., Teknopark İstanbul, 34906, Pendik/İstanbul, Türkiye, arca.bozkurt@iradets.com*
[h] *IRADETS A.Ş., Teknopark İstanbul, 34906, Pendik/İstanbul, Türkiye, deniz.dolek@iradets.com*
[i] *Istituto Nazionale di Fisica Nucleare (INFN), Sezione di Perugia, Via A.Pascoli snc, 06123, Perugia, Italy, lucia.salvi@pg.infn.it*
[j] *BEAMIDE s.r.l., Via Campo di Marte 4/o, 06124, Perugia, Italy, ahmed.imam.shah@beamide.com*
[k] *BEAMIDE s.r.l., Via Campo di Marte 4/o, 06124, Perugia, Italy, nora.ciccarella@beamide.com*
[l] *IRADETS A.Ş., Teknopark İstanbul, 34906, Pendik/İstanbul, Türkiye, yakup.bakis@iradets.com*
[m] *BEAMIDE s.r.l., Via Campo di Marte 4/o, 06124, Perugia, Italy, stefano.gigli@beamide.com*
\* *Corresponding author*

## Abstract

Matter-RADiation interaction SIMulations (MRADSIM®) is an innovative modular software toolkit developed to simulate the effects of radiation on electronic components, human beings and various materials. It incorporates innovative features aimed at enhancing parametric precision, reducing computational time, and introducing supplementary functions for tailored calculations across diverse applications including the applications required for space missions. Notably, MRADSIM is distinguished as the pioneering simulation toolkit to integrate cutting-edge Artificial Intelligence and Machine Learning (AI/ML) algorithms, with the primary objective of effectively recognizing potential radiation-induced issues and facilitating the implementation of mitigation strategies to avert catastrophic failures in mission-critical systems, whether terrestrial or space-based. The distinctive attributes of MRADSIM, coupled with its early adoption by esteemed researchers from the National Institute for Nuclear Physics of Italy (INFN), significantly contribute to the toolkit's added value.

**Acronyms/Abbreviations**

Matter-RADiation interaction SIMulations (MRADSIM), Radiation Hardness Assurance (RHA), Device Under Test (DUT), Artificial Intelligence (AI), Machine Learning (ML), Commercial Off-The-Shelf (COTS), National Institute for Nuclear Physics of Italy (INFN), Standard for the Exchange of Product Data (STEP), Computer-aided design (CAD), Geometry Description Markup Language (GDML), extensible markup language (XML), Graphical User Interface (GUI), Initial Graphics Exchange Specification (IGES), Standard Triangle Language (STL), Virtual Reality Modeling Language (VRML), linear energy transfer (LET), non-ionizing energy loss (NIEL).

## 1. Introduction

Ionizing radiation refers to radiation with enough energy to remove tightly bound electrons from atoms, thereby creating ions [1]. This process can have significant adverse effects on materials, systems, and humans. Ionizing radiation can cause various types of damage to materials, including structural damage by displacing atoms in the lattice of crystalline materials, which leads to defects and weakening of the material structure. In metals, particularly in nuclear reactors, radiation can cause embrittlement, making the metal more brittle and prone to cracking. Some materials swell or change dimensions due to the accumulation of gas bubbles (from transmutation processes) or changes in the crystal structure. Polymers and organic materials can suffer from cross-linking or the breaking of bonds, leading to degradation of mechanical properties and performance.

Ionizing radiation can cause several detrimental effects on electronic systems [2], which is a significant concern in areas like space exploration, nuclear power, and high-energy physics. This along with neutral particles poses significant health risks to humans and other living organisms. The severity of its effects depends on the dose, type of radiation, and duration of exposure.

As previously established, the impact of radiation on humans, materials, and electronic systems can have serious consequences in both short-term and long-term operations, whether in space or on Earth [3]. The continual reduction in the size of semiconductor-based electronics, along with the significant increase in the density of active components, has made electronics increasingly vulnerable to radiation effects. It is crucial to thoroughly study the effects of radiation not only on electronics but also on materials and living organisms at all stages of a project or mission.

Moreover, as the number of private companies launching systems into space continues to grow due to reduced launch costs [4], there is a need to address the challenges associated with using Commercial Off-The-Shelf (COTS) components, which come with inherent radiation risks.

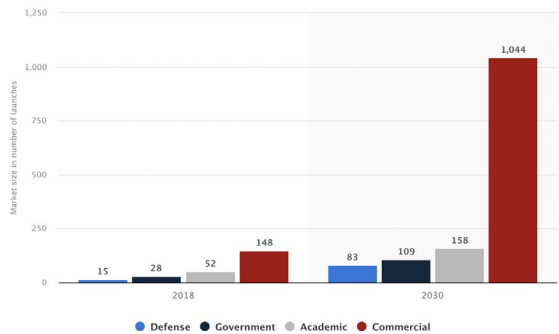

Fig. 1. Market size in number of launches [5].

In the near future, there will be numerous Earth orbit missions and interplanetary explorations to the Moon and Mars, both manned and unmanned. It is essential to conduct comprehensive simulations and on-Earth testing to assess the impact of radiation on humans and instruments intended for operation in these environments. The use of avionics at high altitudes presents similar challenges and requires careful consideration of radiation effects. Furthermore, it is important to preemptively study the effects of accidental or deliberate nuclear explosions on Earth and in space in order to implement appropriate pre- and post-event mitigation measures.

One of the most relevant outcomes from Radiation Hardness Assurance (RHA) studies is the simulation results of radiation effects on targeted devices, materials, or organisms (Devices Under Test, or DUTs). A detailed and realistic simulation should use the actual and final geometry of the mission or project hosting the DUTs and precise knowledge of the local radiation environment. The radiation transport simulation would then proceed particle by particle, step by step, from the source through the geometry to the DUT, or from the DUT to the source, depending on the calculation method determined for the simulation. The radiation analysis parameters and quantities to be simulated will vary based on the mission or project geometry, the DUT structure, and the radiation source(s) being considered. The simulation must accurately model real-life events with a minimal margin of error in analyzed quantities while keeping simulation execution time reasonably short. Additionally, the simulation should feature a user-friendly interface, making it accessible even to non-experts.

*1.1 Problems to Address*

The report from EuroConsult showed that the number of satellites launched from 2009 to 2019 is 2300 with an average of 230 satellites per year [4]. Furthermore, the same report indicates that anticipation is the yearly satellite projects will quadruple from 2019 to 2029. Another report from Statista Research Department indicates a huge market size for the space sector [5] as shown in Fig. 1. Users of radiation simulation applications come from diverse markets, each with a growing number of projects. These projects are becoming more intricate and are being applied in new fields. Consequently, the development of simulation toolkits requires more efficient computing algorithms and systems. It's preferable for these systems to seamlessly integrate with customers' existing information technology (IT) infrastructures. Additionally, the applications must also facilitate users with user-friendly coding technologies, making it possible for a wider pool of potential users to independently manage the underlying complexities. This will extend the use of these technologies to a broader range of applications.

Radiation effects simulation tools are sophisticated software applications. They are specifically created to model and predict the impact of ionizing radiation on different materials, electronic components, systems, and biological entities. These simulations play a critical role in designing and validating products and systems in environments where radiation exposure is a concern. Such environments include space missions, nuclear facilities, and medical treatments involving radiation.

In short, radiation effects simulation tools, like MRADSIM, can be used in different sectors, particularly

for:

- Radiation and Material Interaction Modeling: It simulates how radiation affects different materials, potentially causing changes in structural integrity, electrical properties, or optical characteristics.

- Electronic Component Analysis: Typical effects on electronics include:

  - Single-Event Effects (SEEs): Radiation can cause a single ionizing event that leads to errors in microprocessors, memory, and other electronics [6]. Examples include single-event upsets (SEUs), which can flip bits in memory, and single-event latch-ups (SELs), which can lead to short circuits.
  - Total Ionizing Dose (TID): Accumulation of ionizing radiation over time can degrade the performance of semiconductor devices by increasing leakage currents and changing threshold voltages [7].
  - Displacement Damage: Radiation can displace atoms in a semiconductor lattice, leading to the creation of trap states that affect carrier lifetimes and mobility.
  - Biological Dose Assessment: It predicts the amount of dose received by living organisms in a harsh radiation environment, critical for long-term manned missions and optimization of shielding materials and designs. It is also relevant for Earth-based applications to calculate the radiation type and amount present in houses (natural radiation from soil), radiation therapy environments, and research facilities (radiation sources or accelerators).
  - Environmental Radiation Mapping near and inside Nuclear Implants or Accidental/Warfare Nuclear Explosions: Once the facility design of a nuclear implant and the source are defined, it is possible to predict the effectiveness of shielding design and map the radiation fields inside and outside the infrastructure. It is also possible to simulate catastrophic events and implement proper mitigation solutions.

*1.2 Challanges*

With the increasing number of nuclear projects on a global scale, as evidenced by the organization of the world's first-ever nuclear summit held in 2024 [8], and the growing interest in space projects by private companies, there is a greater need for fast and reliable simulators for these types of projects. MRADSIM® aims to satisfy the overgrowing demand from the industry for the matter-radiation interaction simulations by providing an easy-to-use, reliable and cost-efficient software, while addressing some other issues to enhance the efficiency of its calculations. The major problems that are encountered while dealing with this kind of intricate simulations and calculations can be listed as follows:

- Complexity of Calculations and Software Toolkits: Open-source toolkits such as Geant4 [9], FLUKA [10], PHITS [11], MCNP, and some commercial tools and software like RSIM [12] and FASTRAD [13] require deep understanding and knowledge of related physics and the software itself. The necessity of setting numerous complex parameters can lead to errors and require extensive time and expertise.

- High Computational Demands: The challenges in simulation and computational demands are significant, especially for smaller organizations with limited computing resources. Balancing the need for accurate simulations with available resources can be difficult, often compromising precision. Therefore, there is a real necessity for solutions that combine innovative calculation algorithms and the ability to leverage all available computer resources, such as utilizing CPU/GPU with multi-thread running.

- Lack of Innovative Approaches: To address these computational demands, new accurate and precise calculation approaches, such as AI and ML, are essential. However, none of the existing open-source or commercial simulation tools currently offer these solutions.

- Lack of Interfaces Between Engineering and Physics Worlds: Another challenge lies in the lack of interfaces between engineering and physics worlds. Engineers often design their apparatus in CAD tools, producing data formats that are not compatible with open-source simulation tools. Manual entry of complex designs is time-consuming and error-prone. There is a need for simulation tools that include fast and precise format conversion modules capable of opening major CAD file formats. Additionally, existing solutions often use modeling material around a target material instead of using it in its final position within a design, such as an entire satellite. None of the conversion tools on the market are currently capable of reading material information from CAD files, requiring manual entry, which is time-consuming and prone to errors.

- Non-intuitive Interfaces: Both open-source and some commercial simulation tools feature non-intuitive interfaces, making them difficult to navigate for users unfamiliar with the software.

- Cost: Cost is also a significant factor, particularly for smaller companies, laboratories, or individual researchers, as some commercial tools are expensive and lack a modular structure and flexible licensing models.

**2. MRADSIM® Software**

In order to address the numerous needs for the wide selection of industries that in need of matter-radiation interaction simulations and calculations, MRADSIM toolkit is being developed with a modular architecture that enable the users, whether they are experts or non-experts, tailoring a specialized toolkit for their own needs.

Table 1. Feature list of ConverterFree and ConverterPro.

| ConverterFree Features |
|---|
| Interactive geometry view |
| STEP-to-GDML conversion |
| Adding new material definitions to the database |
| **Added Features for ConverterPro** |
| ≥ x20 faster STEP-to-GDML conversion |
| Large STEP files (≥ 30 Mb) |
| STL, IGES, VRML file type support |
| Improved compatibility with latest STEP file standards |
| Dedicated GDML viewer |
| Overlap check feature |
| Large material database |
| Material assignment on multiple volumes |
| Create and combine simple shapes with preview |
| Add simple shapes to converted STEP file |
| Single piece, multiple pieces and total mass calculation |
| Size measurement tool |
| Search tool for components |
| Screenshot tool |
| Multithread support |
| Multilanguage support (ENG, TR, IT, CN) |
| Technical Support in 48 hours |
| Multiple user/seat licensing option |

*2.1 MRADSIM®-ConverterFree*

This is the first milestone of the MRADSIM project that was completed in 2022. It is a free STEP to GDML format converter available to research institutes and universities for unlimited use for non-commercial purposes.

STEP files are commonly used in computer-aided design (CAD) and in 3D printing to hold three-dimensional model data for a wide variety of design tasks. The Geometry Description Markup Language (GDML) is a specialized XML-based language developed as an application-independent persistent format for defining the detector geometries. The format conversion is required when a user wants to import a geometry created by a CAD into physics simulation tools (i.e. Geant4, GeantV, AdePT, Celeritas etc,.) that read only the GDML format.

MRADSIM-ConverterFree has been benchmarked against other free and commercial products of its kind. The results are published and showed significant performance improvements over other similar tools and software [14]. It is already in use in major research institutes in the world such as NASA, ESA, MIT-Boston, and CERN and universities all over the world in 140 copies in institutions from 26 countries. The current version has limitations on volume size and the number of volumes it can convert.

*2.2 MRADSIM®-ConverterPro*

The Pro version offers more features compared to the free version. Table 1 outlines the additional functions of the Pro version in comparison to the free version of the Converter, including fast opening and conversion of very large STEP files (over 20 times faster than a commercial product), compatibility with all STEP file standards, partial and total mass calculations, a large material database with the ability to define new materials, geometry overlap checks for accurate Geant4 simulations [9], the capability to create user-defined geometries and integrate them with imported STEP files, and essential support for fast shielding optimization investigations.

*2.3 MRADSIM®-Basic*

The MRADSIM-Basic version integrates ConverterPro and specifically focuses on detailed dose analysis on target volumes. It employs both "reverse" and "forward" Monte Carlo simulation methods to expedite the simulation time while accounting for electromagnetic physics. This version offers a 3D dose depth-curve profile and projections on various view axes through the use of AI algorithms.

*2.4 MRADSIM®-Space*

The majority of space missions need to consider three sources of particle radiation:

1. The Earth's trapped radiation belts, which are time-dependent and influenced by conditions in near-Earth interplanetary space.

2. The galactic cosmic ray (GCR) background, which varies slowly over the solar cycle and is anti-correlated with solar activity, as the increased in-fluence of the hemispheric magnetic field results in greater particle attenuation at solar maximum (the impact is greater for low energy radiation).

3. The solar energetic particle (SEP) population, which is sporadic and depends on solar events such as shocks driven by fast and wide coronal mass ejections (CMEs) with a greater frequency of solar particle events (SPEs) seen during solar maximum as opposed to solar minimum.

MRADSIM-Space includes MRADSIM-Basic and calculates quantities such as energy deposition, total ionizing dose, fluence on DUT surface, path length analysis, linear energy transfer (LET), non-ionizing energy loss (NIEL) and charging.

*2.5 MRADSIM®-Earth*

For earth-based applications, MRADSIM-Earth will have a modular structure with pluggable modules related to radiation source type, strength, and position definition for various application fields such as effect of nuclear energy production or explosion, cosmic radiation on Earth's surface, high atmosphere radiation, radiation at accelerators, and radiation in radiotherapy rooms. This version will also evaluate effects at DUT surfaces or inside their fine structures.

Table 2. The file formats and versions that MRADSIM supports.

| File Format | Extension | Version |
| --- | --- | --- |
| STEP | .step, .stp | AP203, AP209, AP214, AP214e3, AP214, AP214e1 |
| STL | .stl | ASCII/Binary |
| IGES | .iges, .igs | IGES 5.3 |
| VRML | .wrl 1 | VRML 2.0 |

## 3. MRADSIM Toolkit – Major Features and Methodology

*3.1 Reading Multiple CAD Files and Conversion to GDML*

MRADSIM Toolkit has a unique user-friendly feature of reading multiple types of CAD files and converting them into GDML. MRADSIM has the capability of read STEP files and convert them to GDML[14]. In addition to the STEP file, MRADSIM can also read STL, IGES and VRML file types and convert to GDML file. Table 2 is a verification table that shows the types of files that MRADSIM Toolkit is able to read and convert to GDML along with the file versions. MRADSIM leverages the OpenCascade libraries to implement the reading and conversion of these files. This makes it an excellent solution for reading CAD files in order to run matter-radiation interaction simulations. The development of MRADSIM continues to increase the supported file types.

While testing the overlapping feature (Section 3.5), the software underwent comprehensive testing to assess its performance in reading STEP files of various sizes and converting to GDML files (Fig. 5).

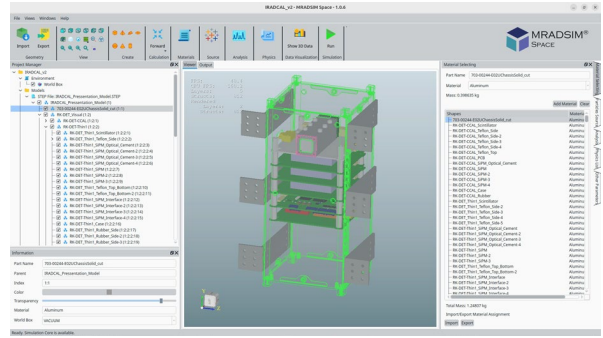

Fig. 2. Graphical User Interface of MRADSIM.

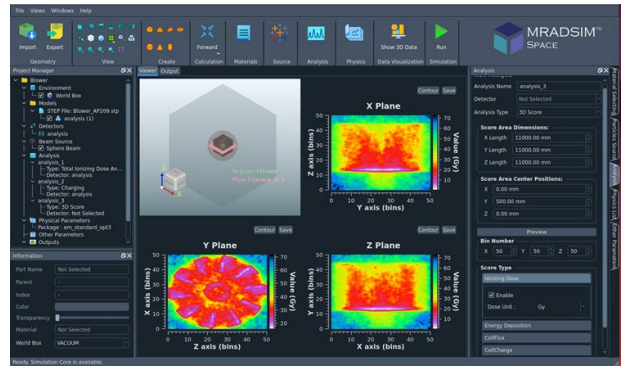

Fig. 3. 3D score results shown for a project in MRADSIM GUI.

*3.2 User-Friendly GUI*

The GUI for MRADSIM Toolkit was created using C++ using QT5 [15] framework [14]. Open CASCADE library is used for the 3D rendering and 3D GUI. The simple yet efficient design of the GUI makes this toolkit very user-friendly such that the user is able to easily import the CAD files and view the 3D geometry in the viewer. Each part and material is assigned a different color making it easier for the user to understand geometry. GUI clearly displays the project hierarchy, material selecting panel, and other functionalities (Fig. 2, Fig. 3). This allows the users to perform no coding and yet leverage the capabilities of GEANT4 simulations[9] and manipulate the 3D geometries to some extent.

### 3.3 Calculation Methods and Physics Library

MRADSIM provides two computation methods: forward and reverse. The forward method employs the conventional Monte Carlo calculation method to transport particles by moving in directions determined from the source. It collects information during interactions with detectors defined on the structures. In the reverse method, primary particles are traced backwards from the sensitive region of the geometry to the external source surface. This method computes only the tracks contributing to the calculations, making it much faster than the standard forward Monte Carlo method when the sensitive region is small compared to the rest of the geometry and the extended external source. The Reverse Monte Carlo method [16] is particularly efficient, but is only applicable to spectrum energy type and spherical source geometry (Fig. 4). Additionally, it is limited to target analysis without cavities and can only simulate electromagnetic processes.

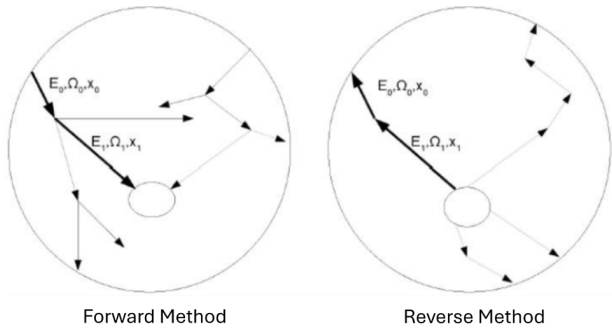

Fig. 4. Forward and Reverse Calculation Methods used by MRADSIM.

MRADSIM provides users with an extensive physics library for comprehensive analysis, enabling the selection of the most appropriate option based on analysis type, energy, and particle. These physics libraries are all frameworks developed by GEANT4. After providing the calculation method and the physics to be used in calculations, the user can define the specifications and the parameters of the radiation source for the simulation.

### 3.4 Adding Geometries and Assigning Materials

One of the important features of MRADSIM is to provide a way to add geometries to the imported project geometry in case of further simulations to expand the data regarding the radiation effects on the DUT, with any additional shielding, so the shielding required for the system or components can be optimized. This provides a quick way to add geometries to the system without the need to alter the CAD file of the system in analysis. Combined with the feature of assigning materials to each component in the system, from a wide selection of materials exists in the modifiable material database that comes with the toolkit.

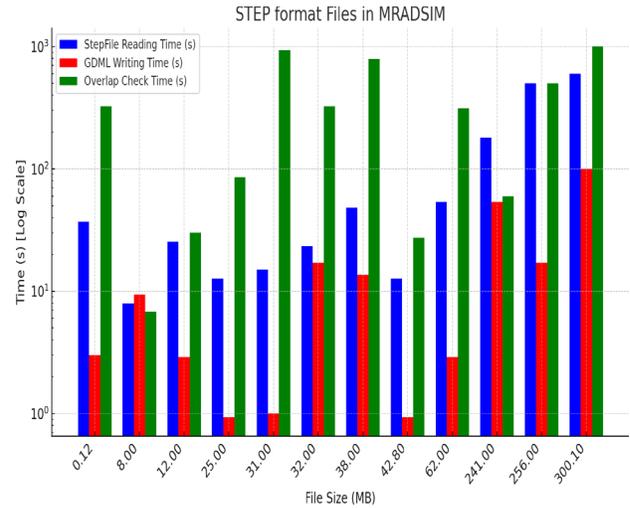

Fig. 5. STEP file reading time, GDML file Writing time and Overlap Check performance (with multithreading) of MRADSIM for STEP files with various file sizes.

### 3.5 Overlap Check

In order to determine if there is any overlapping geometry within the imported design, which can affect the success of the simulation, an overlap check is required prior to initializing a simulation. The duration of overlap checks is contingent upon many different factors, but mostly from the number of individual faces or solids present within the model, which is obviously correlated to the file size, but not completely proportional. Smaller files may contain a greater number of discrete objects, resulting in an increase in pairwise comparisons during the overlap checking process. In contrast, larger files may consist of fewer, larger objects, which reduces the computational burden by limiting the number of required checks.

Precision and Tolerance Parameters: OpenCascade utilizes tolerance values for conducting geometric operations, including overlap detection. The default tolerance can be adjusted depending on the specific requirements of the model in question. Smaller files that demand higher precision are likely to consume greater computational resources, as fine-grained checks for minor gaps prolong processing time. Conversely, larger files with more lenient tolerance settings may complete overlap checks more efficiently, as the algorithm requires less exactitude.

Geometric Complexity: Files with smaller sizes may feature geometrically complex structures with intricate details, such as curved surfaces, sharp edges, and densely packed small features. In such instances, OpenCascade's overlap detection algorithm must perform significantly more computationally intensive calculations, leading to longer processing times despite the smaller file size. Conversely, larger files, which may contain simpler geometries, tend to undergo the overlap-checking process more rapidly.

CAD File Format Specifics: Certain variations in STEP file formats, such as AP209, store geometric data in more complex ways. Even smaller files may experience delays during overlap checks if OpenCascade ([17]) needs to convert or interpret these complex internal representations. For example, although one file (256MB, AP242e1 format) in our study was smaller than the other one, it contained multiple geometric overlaps, leading to prolonged processing times. We utilized multithreaded overlap checks within MRADSIM, with a default tolerance value of $1\times10^{-7}$ mm, to handle these computational challenges more effectively.

Fig. 5 displays the outcomes of comprehensive testing conducted to assess the effectiveness of the multithreading functionality of MRADSIM, which is utilized for overlapping checks, while Fig. 6 shows the success of MRADSIM's multithreading feature for the reduction in execution time for overlap checking.

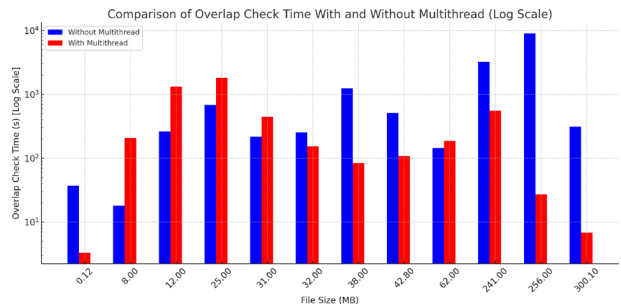

Fig. 6. STEP file reading time, GDML file Writing time and Overlap Check performance (with multithreading) of MRADSIM for STEP files with various file sizes.

### 3.6 Artificial Intelligence & Machine Learning

MRADSIM uses GEANT4 for simulating the dose distributions, dose depositions and energy distributions in the 3D geometries. MRADSIM and GEANT 4 have configuration capabilities where a user can define the number of events and a time along with type of particle source. So in this case when a user defines high number of events such as $10^9$ it takes a lot of time to run such simulations. These type of simulations on large geometries can take up to a couple of days. This creates a lot of hurdle in the field of science and increases the latency of the research. In order to solve this problem, a machine learning model is proposed, Random Forest Regressor[18].

#### 3.6.1 Dataset

To train ML models, a dataset is needed where we have Monte Carlo dose deposit simulations for lower number of events and higher number of events for different geometries. This type of benchmark dataset is not present so a dataset was generated to train these models. The dataset was generated using MRADSIM Toolkit. For each geometry, the MRADSIM toolkit voxelates the geometry in a 50 x 50 x 50 grid. Then the simulation is run for $10^6$ and $10^9$ events. The mission duration for the exposure is set to 1 year. The results of the simulation provide us total dose deposit, number of particle collisions and density information of the geometry according to these voxels. This dataset was generated for four geometries. The models were trained and validated on three geometries and tested on the fourth. The test geometry was left out of the training.

#### 3.6.2 Random Forest Regressor Model

The random forest is an ensemble learning model used for various tasks. The random forest regressors are decision trees that do not overfit on the dataset. For regression tasks, the mean or average prediction of the trees is returned. We train the model by using data from each voxel.

The features used in the training are total dose values for $10^6$ number of events, number of particle entries and density. The output of the model is the total dose values for $10^9$ events.

The random forest model was trained with 2000 estimators, and max depth of 12 branches.

#### 3.6.3 Data Pre-processing

Data pre-processing is really important step for any machine learning algorithm. The first step in data preprocessing is normalization of the dose values. The dose values are normalized as we have data in order of magnitude of $10^8$ Gy. The data needs to be normalized to make sure we have numerical stability and faster convergence that in effect helps in improved model performance. After normalizing, the empty voxels were also removed in order to remove noise from the data.

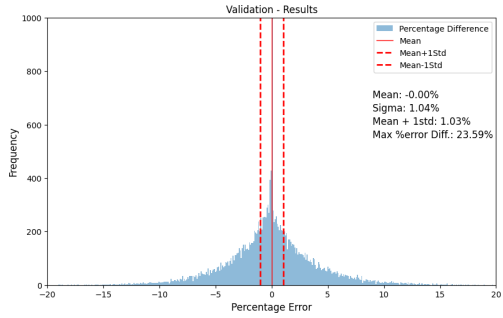

(a) Results from model validated on the 20% of the dataset

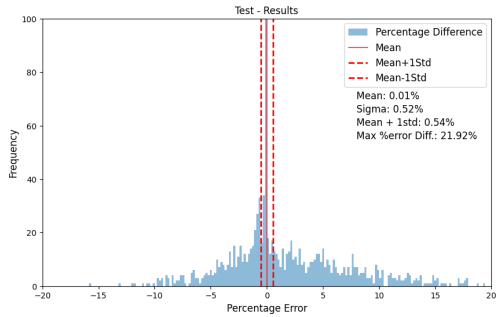

(b) 3D Dose difference plot

Fig. 7. Results from model tested with unknown geometry

*3.6.4 Results*

The model was tested on a test geometry in order to test the model's capability to predict dose for unknown geometries. The criteria used as performance metrics is percentage error displayed in Equation 1.

$$\text{Percentage Error} = 100 \times \frac{Dose_{predicted} - Dose_{original}}{\max(Dose_{original})} \quad [1]$$

The histogram in Figure 7 shows the percentage error for the tested geometry, the y-axis represents the number of voxels and the x-axis represents the percentage error.

The maximum error percentage reported by this model on the test geometry is 21.98% with a mean of 0.01% and standard deviation of 0.52%. This model deals with outliers and anomalies very well. The on average error percentage is good as well as the upper bound of the error percentage that is 21.98%. This model shows good results but are subjected to improve as we proceed with the dataset generation and train the model on more data.

## 4. Toolkit Performance Validation

A benchmarking between MRADSIM-Space and pure Geant4 using GRAS5[19] is made for the same fluxes of trapped electrons, protons, solar protons and heavy ions.

By using the same geometry and same electron spectrum, the comparison results are given in Table 2. The differences between the two simulation results for each calculated quantity, their errors and the differences in terms of the number of sigmas (Eq. (2)) are shown. The results are also consistent for other source particle fluxes (Fig. 8).

$$n_\sigma = (V_m - V_{g4}) / \sqrt{(\sigma_m)^2 - (\sigma_{g4})^2} \quad [2]$$

| | | TRAPPED PROTONS | | | | |
| | | MRADSIM | | GRAS 5 | | |
| Units | | Value | Error | Value | Error | Delta [sigma] |
|---|---|---|---|---|---|---|
| | # events | 4E+08 | - | 400000000 | - | - |
| s | Total time | 56343,5 | - | 87694,2 | - | - |
| us | Time per event | 140,859 | - | 219,2355 | - | - |
| e | Quartz - Charging | 5,53E+09 | 9,22E+08 | 5,68E+09 | 9,34E+08 | -0,117041 |
| MeV | Quartz - Edep | 1,03E+09 | 1,86E+08 | 1,05E+09 | 1,90E+08 | -0,085438 |
| mSv | Quartz - EqDoseQF | 147,723 | 27,243 | 148,522 | 26,4851 | -0,021029 |
| mSv | Quartz - EqDoseRWF | 111,869 | 20,2141 | 114,338 | 20,6702 | -0,085399 |
| particles/cm2 | Quartz - Fluence - total | 7,22E+08 | 1,20E+08 | 7,42E+08 | 1,22E+08 | -0,117043 |
| particles/cm2 | Quartz - Fluence - electron | 0 | 0 | 0 | 0 | - |
| particles/cm2 | Quartz - Fluence gamma | 0 | 0 | 0 | 0 | - |
| particles/cm2 | Quartz - Fluence - proton | 7,22E+08 | 1,20E+08 | 7,42E+08 | 1,22E+08 | -0,117043 |
| MeV/cm | Quartz - LET | 8,07E+19 | 1,79E+18 | 8,10E+19 | 1,76E+18 | -0,119525 |
| 95MeVmb | Quartz - NIEL | 8,73E+11 | 1,51E+11 | 8,70E+11 | 1,48E+11 | 0,016247 |
| mm | Quartz - Path | 1,03E+07 | 1,86E+06 | 1,08E+07 | 2,07E+06 | -0,173949 |
| rad | Quartz - TID | 0,55934 | 0,101071 | 0,571691 | 0,103351 | -0,085419 |
| e | Silicon - Charging | 3,53E+09 | 7,37E+08 | 3,38E+09 | 7,20E+08 | 0,149072 |
| MeV | Silicon - Edep | 6,88E+08 | 1,59E+08 | 5,79E+08 | 1,49E+08 | 0,499295 |
| mSv | Silicon - EqDoseQF | 380,207 | 86,8132 | 296,294 | 74,4485 | 0,733737 |
| mSv | Silicon - EqDoseRWF | 277,662 | 64,0829 | 233,751 | 60,2313 | 0,499297 |
| particles/cm2 | Silicon - Fluence - total | 3,29E+08 | 6,87E+07 | 3,15E+08 | 6,72E+07 | 0,149071 |
| particles/cm2 | Silicon - Fluence - electron | 0 | 0 | 0 | 0 | - |
| particles/cm2 | Silicon - Fluence gamma | 0 | 0 | 0 | 0 | - |
| particles/cm2 | Silicon - Fluence - proton | 3,29E+08 | 6,87E+07 | 3,15E+08 | 6,72E+07 | 0,149071 |
| MeV/cm | Silicon - LET | 5,87E+19 | 2,21E+18 | 6,34E+19 | 3,67E+18 | -1,110623 |
| 95MeVmb | Silicon - NIEL | 4,08E+12 | 1,30E+12 | 1,28E+12 | 4,63E+11 | 2,030266 |
| mm | Silicon - Path | 7,62E+06 | 1,79E+06 | 7,18E+06 | 1,89E+06 | 0,171075 |
| rad | Silicon - TID | 1,38831 | 0,320415 | 1,168750 | 0,301157 | 0,499308 |

Fig. 8. The benchmarking results for the same geometry and same trapped protons fluxes. The last column is the difference in the number of sigma deviation between parameters calculated by MRADSIM-Space and pure Geant4 (10.6) + GRAS 5

## 5. A Study for Radiation Shielding with MRADSIM

The purpose of this simulation was to assess the shielding effectiveness of various materials against gamma rays emitted by Co-60 and Cs-137, using MRADSIM software. The configuration used in the simulation is shown in Fig. 9 where the substrate is Polymethylmethacrylate (PMMA). Table 3 lists the materials chosen for the shielding during this study.

The radiation used to test the shielding effects consists of gamma rays with energies of 1.25 MeV (Average energy produced by Co-60) and 0.661 MeV (energy produced by Cs-137). The simulation was executed in an air environment calculating both the dose and energy deposited in the PMMA and in the shielding material, as well as the dose and energy deposited in the PMMA when no shielding is present. The results shown were obtained by simulating one million events.

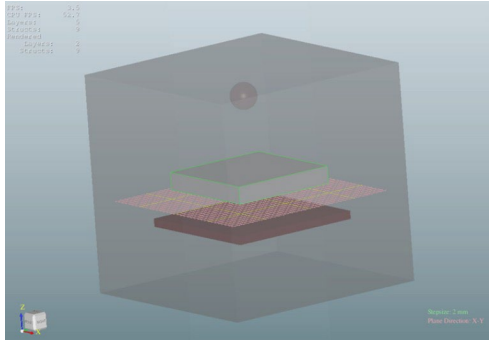

Fig. 9. Geometry used for the radiation shielding study.

Table 3. List of the materials tested for shielding.

| Materials | Density (g/cm$^3$) | Mass (kg) |
|---|---|---|
| Pb | 11.35 | 0.115 |
| Cu | 8.96 | 0.091 |
| In625 | 8.44 | 0.086 |
| In718 | 8.19 | 0.083 |
| 17-4H-SS | 7.75 | 0.079 |
| Al5083 | 2.65 | 0.027 |
| Al2219 | 2.7 | 0.027 |
| B$_4$C | 2.52 | 0.026 |
| SiC | 3.21 | 0.033 |
| Mg | 1.74 | 0.018 |
| H$_2$O | 0.998 | 0.010 |

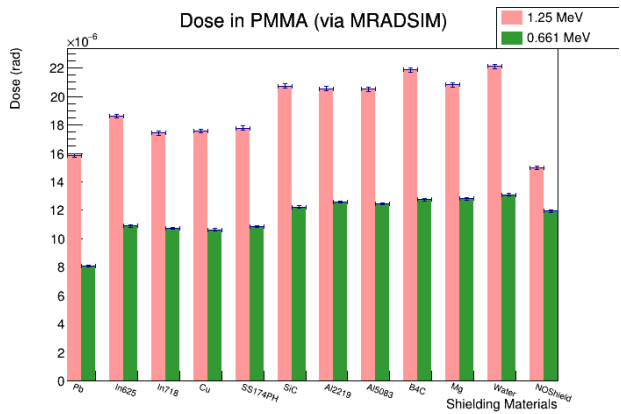

Fig. 10. The total dose absorbed in the PMMA substrate shielded by different shielding materials given by MRADSIM simulation results.

Fig. 10 shows the total dose absorbed by PMMA substrate shielded by the different materials listed from the radiation sources Co-60 and Cs-137. It is worth noting the particular behavior: with certain shielding materials, the energy deposited in PMMA is greater compared to the case where the shielding is not present. It was deemed necessary to compare the observations with a simulation of the same setup in pure Geant4, in order to gain a deeper understanding of the observed phenomenon and to rule out any potential errors in the MRADSIM software.

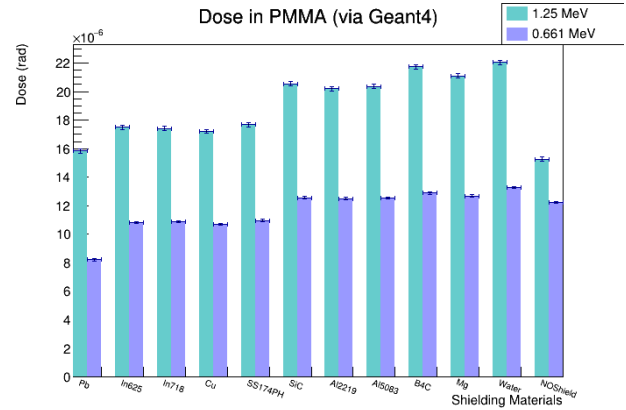

Fig. 11. Total dose absorbed in the PMMA substrate shielded by different materials given by Geant4 simulation results.

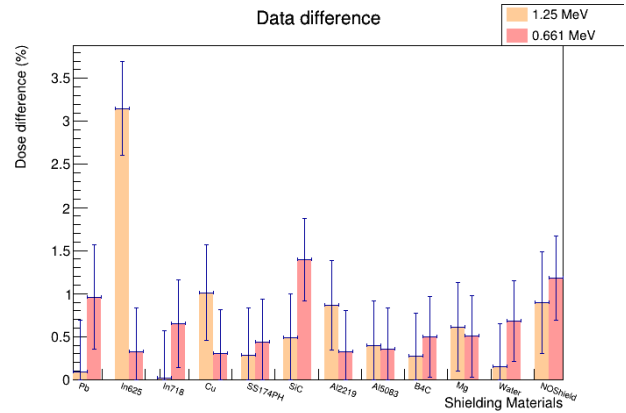

Fig. 12. Relative difference of the total dose absorbed in the PMMA substrate shielded by different materials between Geant4 and MRADSIM simulations.

The geometry used in MRADSIM simulations (Fig. 9) was replicated in Geant4. The material used for the target is G4_PLEXIGLASS from G4NistManager; the shielding materials are the same as listed in Table 3. Using the G4VPrimaryGenerator class, a gamma rays beam was simulated at energies of 1.25 MeV and 0.661 MeV, with the beam oriented along the z-axis. The results in Fig. 11 were obtained by simulating one million events using Geant4. Fig. 12 shows the relative difference between Geant4 and MRADSIM simulation results.

The results obtained with Geant4 are in good agreement with those obtained using MRADSIM. The only case for which the difference is over 3% is for Inconel625 shielding for Co60 simulation, but this can be explained by some composite material definition difference between the two simulations. All other relative differences are below 1.5%, which validates the results produced by MRADSIM.

# 6. Conclusions and Future Works

The flexible design of MRADSIM enables easy customization for different sets of applications, empowering users to select the version that suits their specific testing requirements and analyses. The cross-platform architectural design approach of the application is one of the most innovative aspects of MRADSIM. As a cutting-edge attribute, significant emphasis has been given to developing methods and algorithms for equipping the MRADSIM with innovative AI/ML tools in order to meet the growing and ever-evolving demands of a wide range of industries and emerging applications that require matter-radiation simulations. By utilizing AI/ML, it becomes possible to simulate highly complex projects that are subjected to harsh radiation environments without the need for extensive computing resources or prolonged execution times. Multi-threaded overlap checking in MRADSIM significantly improves performance for large and complex STEP files by distributing tasks across multiple CPU cores. It can drastically reduce processing times, especially for models over 100 MB. Overall, multithreading is ideal for optimizing performance in complex or resource-intensive scenarios, and it is a critical improvement for achieving accurate simulation results. In addition to STEP files, MRADSIM also supports STL, IGES, and VRML formats, ensuring compatibility with a wide range of geometry types. These formats allow for flexible input options while benefiting from the same advanced processing and optimization features. The ML model trained on data collected from MRADSIM, powered by GEANT4, demonstrated remarkable capabilities in generating accurate deposited dose and energy distributions while significantly reducing computational time compared to traditional simulations. However, in the future, the model will be improved with more data and training. Finally, the simulation results done both in Geant4 and MRADSIM for the same geometry and parameters for the total dose accumulation on the PMMA substrate shielded by a set of materials, validate that the MRADSIM simulations are accurate and reliable.

# Acknowledgements

The authors acknowledge the support by both INFN CNTT and TUBITAK during the development of MRADSIM software. Additionally, the authors thank all staff in IRADETS A.S. and BEAMIDE s.r.l. for their continuous and crucial support.

# References

[1] W. Friedberg and K. Copeland, "Ionizing radiation in earth's atmosphere and in space near earth," p. 32, May 2011.

[2] S. Gaul, N. van Vonno, S. Voldman, and W. Morris, *Integrated Circuit Design for Radiation Environments*. Wiley, 2019, pp. 333–343, ISBN: 9781119966340. [Online]. Available: https://books.google.com.tr/books?id=wwazDwAAQBAJ.

[3] J. Talapko *et al.*, "Health effects of ionizing radiation on the human body," *Medicina*, vol. 60, no. 4, 2024, ISSN: 1648-9144. DOI: 10.3390/medicina60040653. [Online]. Available: https://www.mdpi.com/1648-9144/60/4/653.

[4] EuroConsult. "Satellites to be built & launched by 2028: A complete analysis & forecast of satellite manufacturing & launch services." (accessed 2019). (2019), [Online]. Available: https://www.euroconsult-ec.com/research/WS319_free_extract_2019.pdf.

[5] S. R. Department. "Ssmall satellites market size by sector - worldwide 2018 to 2030." (accessed 2024). (2023), [Online]. Available: https://www.statista.com/statistics/1086625/global-small-sat-market-applications/.

[6] D. M. Fleetwood, P. S. Winokur, and P. E. Dodd, "An overview of radiation effects on electronics in the space telecommunications environment," *Microelectronics Reliability*, vol. 40, no. 1, pp. 17–26, 2000, ISSN: 0026-2714. DOI: https://doi.org/10.1016/S0026-2714(99)00225-5. [Online]. Available: https://www.sciencedirect.com/science/article/pii/S0026271499002255.

[7] D. M. Fleetwood, "Total ionizing dose effects on mos and bipolar devices in the natural space radiation environment," Dec. 1998. DOI: 10.2172/10160345. [Online]. Available: https://www.osti.gov/biblio/10160345.

[8] IAEA. "A turning point: First ever nuclear energy summit concludes in brussels." (accessed 2024). (Mar. 25, 2024), [Online]. Available: https://www.iaea.org/newscenter/news/a-turning-point-first-ever-nuclear-energy-summit-concludes-in-brussels.

[9] S. Agostinelli *et al.*, "Geant4—a simulation toolkit," *Nuclear Instruments and Methods in Physics Research Section A: Accelerators, Spectrometers, Detectors and Associated Equipment*, vol. 506, no. 3, pp. 250–303, 2003, ISSN: 0168-9002. DOI: https://doi.org/10.1016/S0168-9002(03)01368-8. [Online]. Available: https://www.sciencedirect.com/science/article/pii/S0168900203013688.

[10] CERN. "Fluka." (), [Online]. Available: https://fluka.cern/.

[11] T. S. et al., "Recent improvements of the particle and heavy ion transport code system – phits version 3.33," *Journal of Nuclear Science and Technology*, vol. 61, no. 1, pp. 127–135, 2024. DOI: 10.1080/00223131.2023.2275736. eprint: https://doi.org/10.1080/00223131.2023.2275736. [Online]. Available: https://doi.org/10.1080/00223131.2023.2275736.

[12] Tech-X. "Rsim - radiation simulation." (accessed 2024). (2024), [Online]. Available: https://txcorp.com/rsim/.

[13] B. Jun, B. X. Zhu, L. M. Martinez-Sierra, and I. Jun, "Intercomparison of ionizing doses from space shielding analyses using mcnp, geant4, fastrad, and novice," *IEEE Transactions on Nuclear Science*, vol. 67, no. 7, pp. 1629–1636, 2020. DOI: 10.1109/TNS.2020.2979657.

[14] A. B. Alpat, A. Coban, H. Kaya, and G. Bartolini, "Mradsim-converter: A new software for step to gdml conversion," *Computer Physics Communications*, vol. 286, p. 108 688, 2023. DOI: https://doi.org/10.1016/j.cpc.2023.108688.

[15] Q. Group. "Qt5." (), [Online]. Available: https://www.qt.io/.

[16] L. Desorgher, F. Lei, and G. Santin, "Implementation of the reverse/adjoint Monte Carlo method into Geant4," *Nucl. Instrum. Meth. A*, vol. 621, pp. 247–257, 2001. DOI: 10.1016/j.nima.2010.06.001.

[17] O. C. Technology. "Opencascade." (), [Online]. Available: https://dev.opencascade.org/project/mradsim.

[18] A. Liaw and M. Wiener, "Classification and regression by randomforest," *Forest*, vol. 23, Nov. 2001.

[19] ESA. "Gras." (), [Online]. Available: https://space-env.esa.int/software-tools/gras/.